\documentclass[letterpaper]{article}
\usepackage{graphicx}
\usepackage{hyperref}
\hypersetup{
    colorlinks=blue,
    linkcolor=blue,
    filecolor=blue,      
    urlcolor=blue,
    }
\usepackage{geometry}
\usepackage{algorithm}
\usepackage{algpseudocode}

\usepackage{caption}

\usepackage[longnamesfirst,sort]{natbib}
\bibpunct[, ]{(}{)}{;}{a}{,}{,}

\title{Complexity of frequency fluctuations and the interpretive style in the bass viola da gamba}
\author{Igor Lugo\thanks{Corresponding author: \href{igorlugo@crim.unam.mx}{igorlugo@crim.unam.mx}}\\Universidad Nacional Aut\'{o}noma de M\'{e}xico (UNAM)\\Centro Regional de Investigaciones Multidisciplinarias (CRIM)\vspace{0.5cm}\\ Martha G. Alatriste-Contreras\\Universidad Nacional Aut\'{o}noma de M\'{e}xico (UNAM)\\ Facultad de Econom\'{i}a (FE) \vspace{0.5cm}\\ Rafael S\'{a}nchez-Guevara \\ Universidad Nacional Aut\'{o}noma de M\'{e}xico (UNAM) \\ Facultad de M\'{u}sica (FaM)}

\date{May 20, 2025}

\begin{document}

\maketitle

\begin{abstract}
Audio signals in a set of musical pieces are modeled as a complex network
for studying the relationship between the complexity of frequency fluctuations and the interpretive style of the bass viola da gamba.
Based on interdisciplinary scientific and music approaches,
we compute the spectral decomposition and translated its frequency components to a network of sounds.
We applied a best fit analysis for identifying the statistical distributions that describe more precisely the behavior of such frequencies and computed the centrality measures and identify cliques for characterizing such a network. Findings suggested statistical regularities in the type of statistical distribution that best describes frequency fluctuations. The centrality measure confirmed the most influential and stable group of sounds in a piece of music, meanwhile the identification of the largest clique indicated functional groups of sounds that interact closely for identifying the emergence of complex frequency fluctuations.
Therefore, by modeling the sound as a complex network, 
we can clearly associate the presence of large-scale statistical regularities with the presence of similar frequency fluctuations related to different musical events played by a same musician.
\end{abstract}

\section{Introduction}
Playing a piece of music is a singular event that shows one of the intrinsic attributes of music: the complexity of frequency fluctuations. For example, even though a musician makes an effort to control the physical aspects of music---i.e., properly tune the musical instruments, to control the string tension by regulating the air-conditioning system in a studio, and to follow strictly the Western musical notation---the generated sounds do not correspond exactly to the fundamental frequencies in Hertz (Hz) associated with the scientific pitch notation of the equal-tempered scale or unequal temperaments \citep{Fineberg2000,Lindley2001,Byrd2007}.
The presence of such fluctuations in music, on the one hand, is related to the common ideas of uniqueness and irreproducibility---a snapshot of sounds---because they capture a sequence of unique sounds in a time frame. 
On the other hand, in a scientific approach, such fluctuations can describe the presence of irreducible and unpredictable sounds in music. Either way, they describe some part of the music as a natural system of sounds. This type of system shows a great diversity of behaviors that cannot be effectively measured and predicted. In other words, this type of system can be associated with the concept of computational irreducibility because to fully understand its dynamics it is required to execute or simulate the entire process based on individual and collective behaviors and observe the trajectories that emerge \citep{Wolfram2002,Rowland2025}. 
In addition, in every particular performance of music the interpretative style of a musician---i.e., individual expression throughout particular musical instruments for communicating author's intentions---can highly contribute to such fluctuations \citep{Perlovsky2017,Solomonovaetal2023}. 
For example, some of the frequency fluctuations can be related to the presence of the immediate musical composition---improvisation---or pre-planned variations in a piece of music.
Therefore, to model such fluctuations and to associate them with the interpretative style are not trivial issues. For the purpose of this study, we use the complex systems and the music theory approaches for exploring  
the relationship between 
the presence of complexity in frequency fluctuations and the interpretative style in a set of pieces of music related to the bass viola da gamba. We propose to model a network of sounds based on the spectrum of audio signals of each piece of music.

The study of music based on complex networks is not new in the literature of musical notation. For example, the work of \citet{Luietal2010} showed how to use a network of music in which nodes and edges were related to musical notes and their co-occurring connections for composing music. Following the same line, the work of \citet{Ferretti2017} proposed to use a complex network approach for modeling musical solos. However, the complexity approach when using sounds related to pieces of music of audio recordings instead of the traditional staff notation has not been explored in detail. For example, the use of an audio signal processing can assist to identify underlying statistical regularities in early music \citep{LugoAlatriste2025} and to generate contemporary music based on simple algorithms based on such statistical regularities \citep{LugoAlatriste2024}. However, one of the main concerns of these studies has been to identify the effects of the interpretative style of musicians when playing their own pieces of music or the music of others. 
Therefore, the scientific study of music as a system of sounds remains an open question because of the presence of such frequency fluctuations and their irreducible human attribute.

Consequently, with regard to our interdisciplinary approach and the irreducible condition of the sound, we aim to explore the most probable answers of the following questions: What type of a network is the network of sound based on complex frequency fluctuations?
Does the modeling of a piece of music as a network of sounds assist to understand frequency fluctuations?
What types of complex network measures can characterize the relationship between the spectrum of audio signals and the irreducibility of the interpretative style? 
To answer these questions, 
we use an explorative data analysis based on statistical physics and music theory \citep{AlbertBarabasi2002,Morrison2010,HolmHudson2016}.
In particular, 
we are interested in identifying the relationship between statistical regularities---best fit statistical distributions---and the properties of the complex network of sounds---centrality and cliques---based on the spectrum of audio signals.
We hypothesize that, at a large-scale, the presence of statistical regularities may distinguish singular interpretative styles. At the same time, at a lower scale, the structure of the network of sounds describes some of the complex frequency fluctuations derived from the interpretative style at a particular moment. 
We expect to find the presence of similar statistical distributions in the set of different pieces of music due to the performance of a same musician, and subtle differences in the complex network measures because of the granular and more precise details captured in the formation of a network of sounds. Therefore, the interpreter imprints particular frequency fluctuations in every piece of music
in which a predictable pattern of sounds coexists with their unpredictable variations. 

One of the main goals of the present study is to contribute to the dialogue and discussion regarding the human creativity and its role in the human arts---e.g., the music---and the role of the automatization of such a creativity by computer algorithms that could reproduce exactly and equivalently in their sophistication the human creativity---the principle of computational equivalence \citep{Wolfram2002,Rowland2025}. In any case, the logic behind human creativity and creativity by computer algorithms is one of the ultimate scientific quests. The next sections are organized as follows: second section covers the literature review, following we present the section of Materials, Methods, Results, and Discussion. 

\section{Literature review}
Following the idea of the effect of the interpretative style in the presence of complex frequency fluctuations, 
we have distinguished two branches of literature related to reducible and irreducible attributes of a system of sounds. In particular, the reducible approach deals with the presence of complex frequency fluctuations by encapsulating sounds in the form of discrete musical notation---e.g., the musical score.
Then, the sound is reduced into a simple and predictable system's behavior. For example, the staff notation is a visual representation of music based on individual and collective symbols for given musical notes.
One of the latest studies in the development of this approach is the work of \cite{BuongiornoNardelli(2020)}. He proposed a method in which the musical structures can be modeled as a complex network for compositional practice.
On the other hand, the irreducible approach models such frequencies as a continuous variable associated with an audio signal measured in Hz based on the equal-tempered scale.
Then, the music can be described as a complex and unpredictable system of individual and collective sounds. For example, in this respect, we specially mention the work of \cite{Busoni1911}. He considered the use of microtones---to extent the current musical system of the twelve-tone equal temperament by adding subdivision of semitones---to increase the range of possible useful sounds in music \citep{LugoAlatriste2024}.
Therefore, both branches of literature emphasize the study of complex frequency fluctuations by using different measures and methods. In this study, we are particularly interested in mentioning the literature associated with the latest approach.

Based on the irreducible approach, we have to mention the work of \cite{KlapuriDavy2006} and \cite{MullerKlapuri2014}. The former addressed some problems related to the automatic music transcription based on different signal processing methods. The latter showed the development of principle ideas and methods for retrieving and analyzing data in music signals. The work of \cite{Downey2016} described a programming-based approach for explaining the process and analysis of digital audio signals. And finally, the study of \cite{LugoAlatriste2025} who showed the use of both cited references in an applied study for displaying statistical regularities of audio signal related to the musical work of Marin Marais. These studies are some examples of the advances of using an interdisciplinary approach in the musical analysis.  

In addition, considering the importance of the human influence on the process of interpreting and performing any piece of music, and therefore contributing with the formation of complex frequency fluctuations,
the work of \cite{Solomonovaetal2023} identified key factors that describe the preparation of the performer. They suggested that studying and knowing the social and individual context of the composer matters for a high quality of the performer's expressiveness. In particular, they suggested that the interpretative style is similar to the actor's performance in which an actor creates a believable and engaging character in a biographical story. In the same line of our research interests about identifying significant differences between interpretative styles by musicians who play the same musical score, \cite{Heroux2018} suggested that the context, musical work and external constrains impacted the strategies used by musicians. Moreover, the work of \cite{Molino2011} recognized the importance of considering the music as an irreducible and singular event. The music as a complex process of sounds between the composer, the performer, and the listener. Finally, it is important to note the fact that not only studies of creativity based on Humanities have identified different and related elements that describe the process of interpretation and performance in artistic domains, but also studies related to social and computer sciences have explored such a creativity process. For example, the work of \cite{LugoAlatriste2024a} used a 2D cellular automata for understanding and replicating aspects of the musical creativity.

To conclude this section, the literature in this respect identifies the complex phenomena of understanding frequency fluctuations and tracing their key elements in a musical context. Even though there are different approaches for studying sounds and their relationships by making music, the goal is to clarify some aspect of the creativity process. The next section, we shall describe the audio records behind our analysis.

\section{Materials}
The audio records are related to a particular repertoire of Rafael Sánchez Guevara. Such set of musical pieces corresponds to different composers across different periods of time. In particular, 14 audio files were analyzed in this study. All of them contain recordings of live performances on the viola da gamba, captured in three different
musical time frames (Table \ref{lpr})

\begin{table}[!ht]
\scriptsize
\centering
\begin{tabular}{|l|c|c|c|}
\hline
{\bf Audio files} & {\bf Date} & {\bf Venue} & {\bf Composer} \\
\hline
\hline
1Preludio\_re\_mayor	& Apr 24, 2020 & Home recording & Carl Friedrich Abel\\
2Arpegaita			& &  &\\
3Cadencia			&  & &\\
4Otra\_cadencia		&  & &\\
5Fuga				&  & &\\
6Cadencia\_pizz		&  & &\\
					\hline
Demachy\_Sarabande\_re\_menor	& Oct 22, 2020 & San Sebastián Mártir Parrish, & Monsieur Demachy,\\
Demachy\_Suite\_Chaconne\_sol	& & Mexico City & Monsieur de Sainte-Colombe\\
Demachy\_Suite\_Gavotte			& &  &\\
Demachy\_suite\_preludio			&  & &\\
Demachy\_Suite\_Sarabande\_sol	&  & &\\
Sainte\_Colombe\_Chaconne		&  & &\\
Sainte\_Colombe\_tres\_preludios	&  & &\\
					\hline
Rafa\_Abel\_Arpegiata\_042020	& Oct 8, 2022 & Recording studio & Carl Friedrich Abel\\
			\hline
\end{tabular}
\caption{\label{lpr}{\bf Recordings of live performances.} }
\end{table}

Table \ref{lpr} shows pieces of music related to three musicians, composers and performers: Monsieur Demachy and Monsieur de Sainte-Colombe, composers from the seventeenth century; and Carl Friedrich Abel, composer from the eighteenth century. All of them were well-known viola da gamba performers at their living times. 

Monsieur Demachy (fl. 1685) and Monsieur de Sainte-Colombe (fl. c. 1680) were two
musicians who share several characteristics and also relevant differences. Their biographic details are very scarce, and both authors were known only by their family name, making it difficult to find precise data about their lives. Both musicians lived in Paris in the second half of the seventeenth century and composed exclusively for the viola da gamba. While Demachy composed only solo pieces, Sainte-Colombe wrote music for one and two violas da gamba. Demachy is accounted as the first composer to have published printed music for viola da gamba in France. In particular, a book containing eight suites, Pièces de violle \citep{Demachy1685}. On the other hand, Sainte-Colombe did not publish his work, but his legacy is preserved only in hand-written notebooks. Those two composers are also pioneers in requiring the use of the seventh string of
the viola da gamba (traditionally a six-string instrument until then), which was an innovation attributed to Sainte-Colombe by Jean Rousseau in his Traité de la viole \citep{Rousseau1975}.

Historically, the decade of 1680 is a landmark in the history of the viola da gamba in France. In that time, there were several of the most important publications. \cite{Demachy1685} published the first collection of music for viola da gamba, meanwhile only one year before Marin Marais published his first book of pieces, Pièces à une et à deux violes, Premier livre, 1686 \citep{Savall2007}. In 1687, two of the main French treatises of technique and performing style were published: Traité de la viole \citep{Rousseau1975}, and Le Sieur Danoville L’art de toucher le dessus et basse de violle. The musical works by Demachy and Sanite-Colombe were innovative and showcased a distinctive composition and playing style commonly referred to as style brisé. Such a style is characterized by irregular and unpredictable melodies, broken chords in leaping lines, and the lack of signature or barlines in some pieces related to preludes \citep{Ledbetter2001}.

About Carl Friedrich Abel (1723-1787), he lived around one-hundred years after Demachy and Sainte-Colombe. He had a successful career in Germany and England as a soloist concert player, promoter and composer. He belonged to a family of musicians in which his father Carl Ferdinand Abel was a colleague with Johann Sebastian Bach in Köthen. When he was young, Abel studied in Leipzig under the guidance of Bach up to the age of 20 years old \citep{Holman2010}. He was an outstanding entrepreneur in partnership with Johann Christian Bach---Johann Sebastian’s youngest son. Together they founded the ``Bach-Abel concert series'' in London, the first project of its genre in the history of Classical music. In addition, they managed a concert hall where they hosted public recitals featuring the most celebrated musicians of the time, probably including five-year old W.A. Mozart.

Abel is considered the last famous virtuoso of the viola da gamba. He composed a large amount of music including sonatas, duos, and solo pieces. The seven compositions included in this study (Table \ref{lpr}) are contained in the Drexel Manuscript, MS 5871 \citep{NYPL2025}. It is a personal notebook, not published during the author’s life, that contains around 25 pieces for unaccompanied viola da gamba \citep{Cyr1987}. Such a manuscript is preserved in \cite{NYPL2025}, and it shows developed instrumental and compositional techniques related to complex polyphonic structures and ``minimalistic'' writings---i.e., the pieces of Fugue and Arpeggiata respectively. Some of such pieces were unfinished or extremely short musical works that were meant to be further works or improvisations.

It is worth mentioning that for all of these recordings (Table \ref{lpr}), the performer plays on a seven-string viola da gamba made by Francis Beaulieu in Montreal, Quebec, in 2012. This instrument is a replica of an original viola da gamba made in 1683 by Michel Colichon. He lived in the same city and time as Demachy and Sainte-Colombe. Nowadays, the original viola is preserved in the Musée de la Musique (Cité de la Musique) in Paris. That original Colichon's viola is the oldest surviving example of a seven-string viola da gamba, and it represents a milestone in the French lutherie tradition. This old instrument is not in playing conditions, but has been thoroughly analyzed and served as a model for a large number of modern reproductions. Its main features are particular resonances and registers. It is the smallest model of that kind of instruments (67 cm body length and 69,5 cm of vibrating string length). It is strung with hand made gut strings: 4 uncovered on the top register, and 3 wire-wound with silver on the bottom. 

Additionally, a relevant issue about music recordings on a particular instrument is that their characteristics depend on a number of factors such as recording and postproduction techniques, equipment, edition, fidelity and acoustic of recording, and playing displays. The presence of these factors must prevent us to equate the analysis of a recording and the analysis of a musical performance \citep{Clarke2006}.

Finally, we used different Python libraries to retrieve, analyze, and plot the data and results. In particular, we use the \textsc{Numpy}, \textsc{Matplotlib}, \textsc{Scipy}, and \textsc{Pandas}. The database and the code will be available in the project name
 \href{https://osf.io/x2qat/?view_only=2ab85b52307b43388da4b52212f2598f}{\emph{The complexity of frequency fluctuations in music}}, Open Science Framework (OSF), for transparency, openness, and reproducibility in science \citep{Noseketal2015}.

\section{Methods}
\begin{figure}[!h]\centering
	\includegraphics[width=0.52\textwidth]{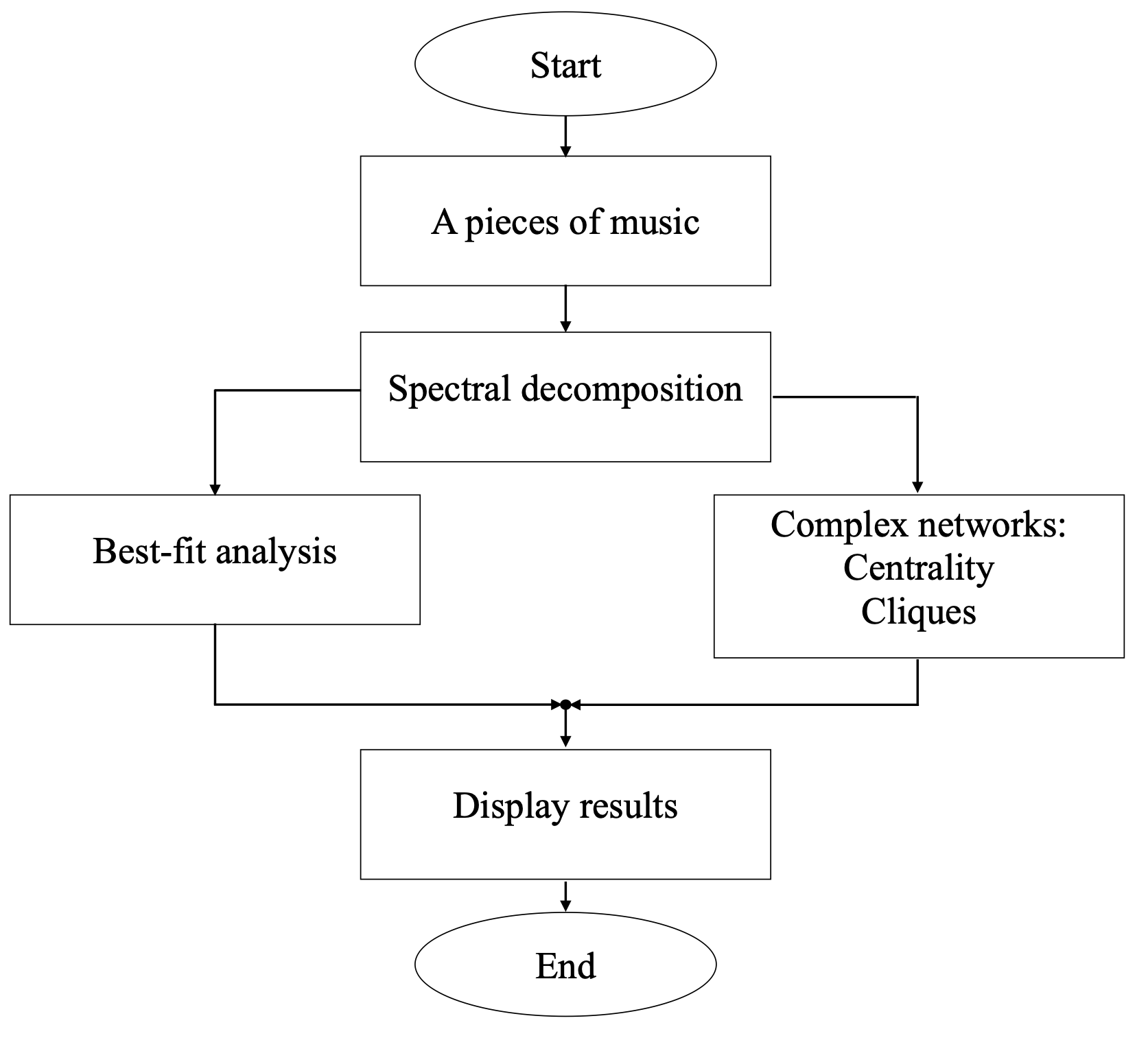}
	{\caption{Flowchart of the method.\label{flowchart}}}
\end{figure}

We are proposing a novel approach for identifying large-scale statistical regularities in a set of digital audio signals and showing their small-scale network variations related to complex frequency fluctuations. In particular, we modeled a network of sounds in which the output of the spectral decomposition is used as input for generating such a network. Therefore, our approach uses audio signals and initiates the data analysis by computing the spectral decomposition (Figure~\ref{flowchart}). The output of this computation generates a sequence of components related to Hz. Next, based on such components, we applied a best fit analysis for identifying the statistical distribution that best describe them. After this, we translate this sequence of Hz to a network configuration in which each node is related to a particular sound expressed as a range of Hz. Consequently, this sound is associated with a pair of particular notes related to the well-known musical system of the twelve-tone equal temperament. Edges between nodes are related to pre- and post-sounds of each current value of the sequence of components. 

Therefore, our method aims to unveil the complex systems attributes related to the presence of similar patterns of complex frequency fluctuations. In the next subsection, we shall describe each of this steps in more detail.

\subsection{Spectral decomposition}
The spectral decomposition is a method that simplifies a waveform in a set of its constituent frequencies to identify approximately their pitches. This method is related to the Fast Fourier Transformation (FFT) algorithm and the Discrete Fourier Transformation (DFT)~\citep{Cooleyetal1965, Press2007,SciPy2025}. In particular, we used the following FFT:

\begin{eqnarray}
\label{eq:1}
	y[k] = \sum_{n=0}^{N-1} e^{-2\pi j \frac{kn}{N}} x[n]
\end{eqnarray}
in which $y[k]$ is the frequency component of a sequence of the audio signal $x[n]$ from $n$ to $N-1$. Therefore, after computing the spectral decomposition for every piece of music, we can generate a list of constituent frequencies in Hz of the waveform. 

\subsection{Best fit analysis}
We used the best fit analysis to identify the statistical distributions that accurately outline each piece of music related to its frequency components. We utilized the Kolmogorov-Smirnov (KS) goodness-of-fit test for comparing each data of frequency components with a set of theoretical distributions \citep{Massey2020}. In particular, based on the work of \cite{LugoMartinezMekler2022} and \cite{LugoAlatristeContreras2022}, we used the following theoretical distributions: normal, log-normal, exponential, Pareto, Gilbrat, power law, and exponentiated Weibull. In the family of univariate distributions, these distributions represent the most general probability distributions that contain special cases and transformations of other distributions~\citep{WilliamMary2025}. In particular, we compute the KS test for each theoretical distribution, then visualize the empirical and theoretical cumulative density functions (CDF). After estimating the statistics and p-values of the KS test and visualizing the fit of CDFs, we can identify which theoretical distributions is the most suitable to describe our data.

\subsection{Complex networks: degree centrality and cliques}
We generated a network of sounds as an undirected graph. Following the notation of \cite{AlbertBarabasi2002}, we can define this graph as a pair of sets $G=(P, E)$ where $P$ is a set of $N$ nodes, $\{P_1, P_2, ..., P_N\}$, and $E$ is a set of edges that connect two nodes, such as $\{P_1, P_2\}$. In our study, edges are unordered pair of nodes, for example $\{P_1, P_2\}$ is the same as  $\{P_2, P_1\}$. Then, each node, $P_i$, of this graph corresponds to each of the constituent frequencies related to the spectral decomposition. In particular, because the value of such frequencies falls between a pair of their closest frequencies associated with the equal-tempered scale, we model each node as a pair of such closest notes,
for example, $P_1 = \{E4, F4\}$ or $P_1 = \{329.63, 349.23\}$ where each of them corresponds to the scientific pitch notation and the Hz associated with such a notation respectively.
Furthermore, edges correspond to a pair of pre- and post-constituent frequencies related to a current value in the list of constituent frequencies. For example, selecting a particular constituent frequency based on its order of appearance in the list of frequency components, we can define a connection between two nodes based on the preceding and current frequency components, and other connection is formed by the current and following frequency components. 
For example, the edge $\{P_1, P_2\}$ can be associated with $\{\{E4, F4\}, \{B3, C3\}\}$ where $P_2 = \{B3, C3\}$.
Therefore, we call this type of network as a network of sounds, which is a collection and unique realization of related sounds that are closely related to each other. Next, we shall compute its degree centrality and find the largest clique in such a network.

The degree centrality computes the fractions of nodes that are connected to a particular node \citep{Hagbergetal2008}. In this case, we are interested in identifying nodes, pair of notes associated with the equal-tempered scale, that show the most number of connections with other nodes. In other words, we expect to identify key notes in the network of sounds that reflect the importance of them with respect to their connection with the preceding and following sounds.

A clique in a network is a subset of nodes in which every node is connected to other node \citep{Hagbergetal2008}. In our case, the largest clique on a network of sounds can show the closely interacting sounds that work together \citep{BronKerbosch1973}. Particularly, sounds that are not commonly related to each other in music theory, but are commonly related in the nature of the frequency fluctuations.

\subsection{An example of application}
In this subsection, we show an implementation of our proposed method. For the sake of simplicity, we use one piece of music of the data set---Demachy\_Suite\_Chaconne\_sol---display a figure that features the main aspects of the data analysis, and we show other figure that encapsulates the degree centrality and the largest clique in a network of sounds. Such figures are intended solely for exemplifying the application of our method. We will not display these figures for each piece of music in the result section.

\begin{figure}[!ht]\centering
	\includegraphics[width=0.6\textwidth]{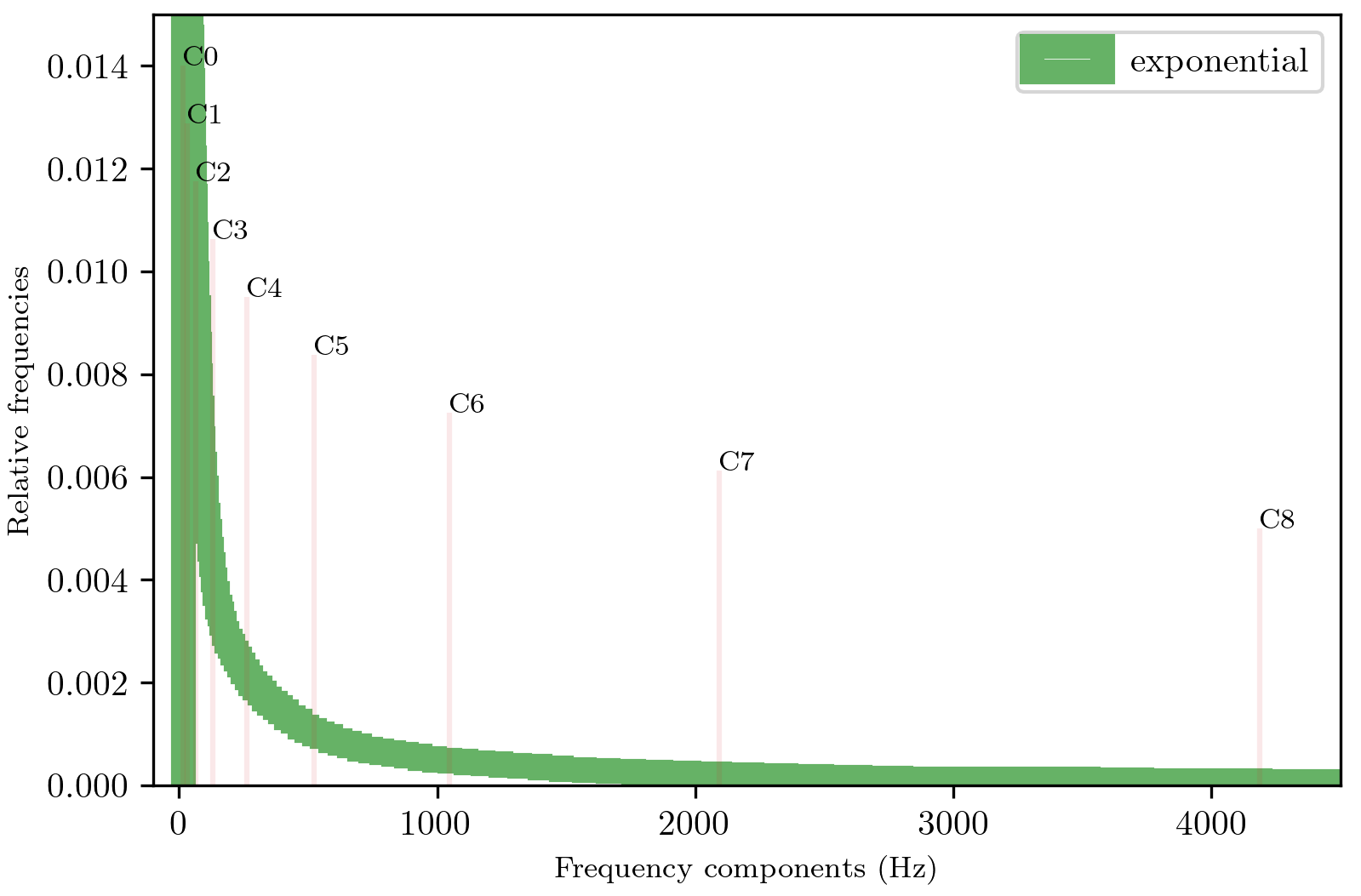}
	{\caption{Best fit statistical distribution. The digital audio is related to the file of ``Demachy\_Suite\_Chaconne\_sol.''
	\label{DemachySuiteChaconnesol}}}
\end{figure}

Figure \ref{DemachySuiteChaconnesol} shows the statistical regularity found in this example of implementation. Based on the method of spectral decomposition, we use the resulted sequence of frequency component as inputs for identifying the statistical distribution that best fits the data.

As we can see in Figure \ref{DemachySuiteChaconnesol} the frequency histogram of the Hz frequency components shows a skewed behavior of data. After using the best fit analysis, we identified the exponential distribution as the best approximation.

\begin{figure}[!ht]\centering
	\includegraphics[width=0.7\textwidth]{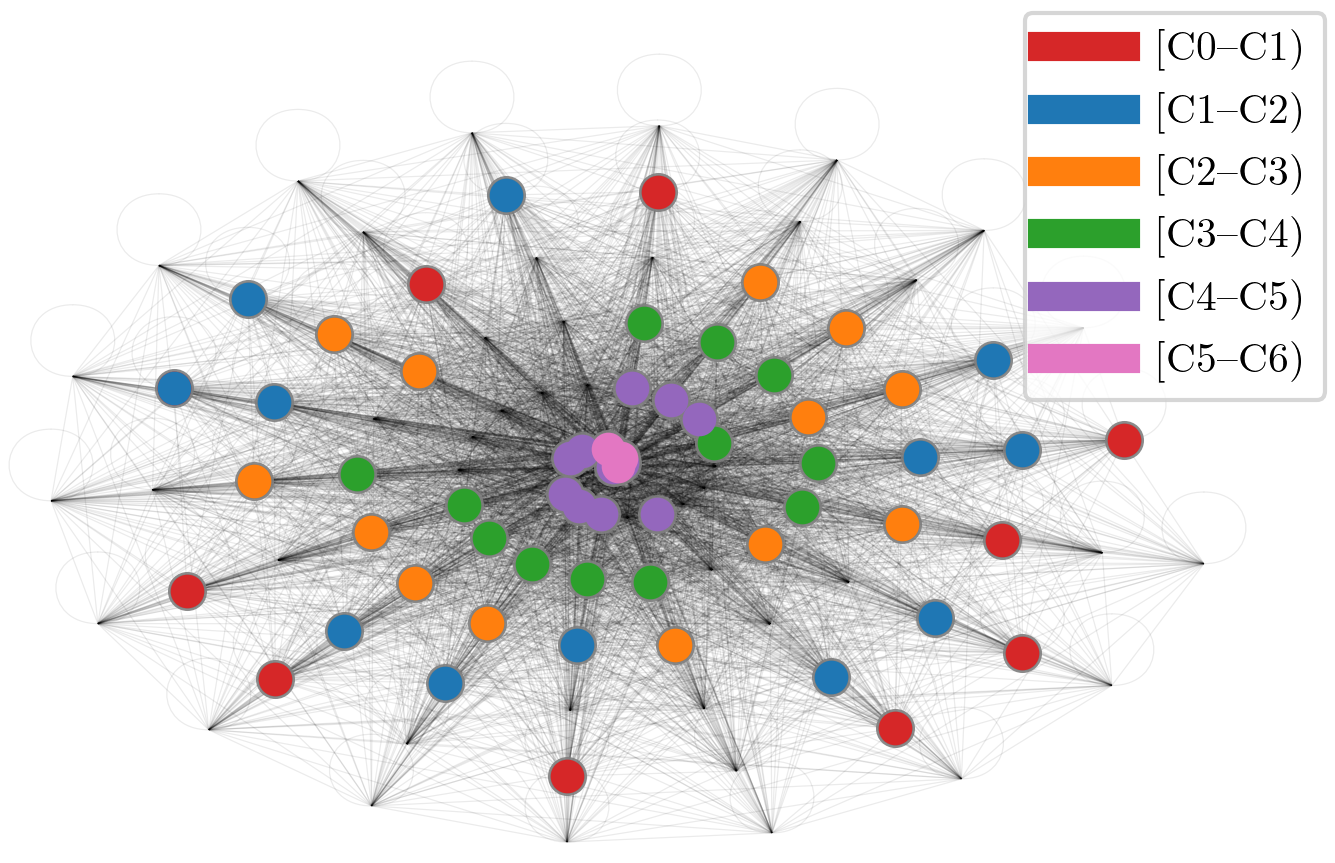}
	{\caption{Network of sounds. Spiral layout where the positions of nodes are related to their degree centrality. Only nodes belonging to the largest clique appear. The color of nodes is related to notes.
	 \label{DemachySuiteChaconnesolNetwork}}}
\end{figure}

In addition to the best fit analysis, we can see in Figure \ref{DemachySuiteChaconnesolNetwork} the network of sounds based on the Hz frequency components of the spectral decomposition. The location of each node is related to its degree centrality in which the higher value is located at the center of the figure, and the subsequent nodes are deviated from the center in a spiral shape. Furthermore, the largest clique is displayed by colors that show the different range of sounds based on the scientific pitch notation of the equal-tempered scale.

We observe in Figure \ref{DemachySuiteChaconnesolNetwork} that the most central nodes are at the center or core of the spiral. These nodes are related to C5-C6 and C4-C5 in the note system.

Therefore, for every musical case investigated in this study, we use the best fit analysis, generate a network of sounds, and compute the degree centrality and the largest cliques. However, we do not show the distribution or the network for each musical piece. We only show numerical results. Furthermore, we compared each musical case with each other for identifying emergent patterns behind the collection of such audio signals. In the section that follows, we present our findings that show our explorative data analysis for identifying the best fit statistical distributions, the degree centrality, and the cliques of each musical case related to our database.

\section{Results}
Based on the idea that in every piece of music we can find the presence of complex frequency fluctuations, the aim of this study was to investigate the relationship between such frequency fluctuations and the interpretative style of a particular musician playing the bass viola da gamba.

The process to study such a relationship is related to the complex systems and the music theory approaches. As mentioned earlier, we used an explorative data analysis for identifying the best fit statistical distributions, the degree centrality, and the cliques of the spectrum of audio signals. Each of them displays consistent information that identifies a singular interpretative style and its particular features related to a network of sounds.

We shall start by showing the results related to the identification of statistical regularities in our set of pieces of music (Table \ref{vbestfit}). As shown in Table \ref{vbestfit}, second column, the statistical distribution that best fits the spectral data in almost all cases is the exponential. The only exception to this is the ``4Otra\_cadencia.'' It showed that the gibrat statistical distribution best fits the spectral data. It is important to remind that this distribution is a special case of the lognormal. Therefore, after doing the best fit analysis, we noticed a clear presence of the exponential distribution just over 92\% of all cases. The third column in Table \ref{vbestfit} shows the estimated parameters in the KS test.

\begin{table}[!ht]
\scriptsize
\centering
\begin{tabular}{|l|c|c|}
\hline
{\bf Piece of Music} & {\bf Best Fit (parameters)} & {\bf KS test(d, p-value)} \\
\hline
\hline
1Preludio\_re\_mayor	& \textbf{exponential} &\\
					& (8.217361091457574e-05,& (0.0003677609528815462, \\
					&29.796745573406135)  &0.9316238664812938)\\
					\hline
2Arpegaita		& \textbf{exponential} &\\
				& (9.399969081655193e-05,  & (0.0003912449672245355, \\
				& 43.87649608115157) & 0.677978790881924)\\
				\hline
3Cadencia		& \textbf{exponential} &\\
				& (3.087383980079031e-05, &(0.0007488242771395148,\\
				& 17.27483455459813) & 0.7974351536960372)\\
				\hline
4Otra\_cadencia 	& \textbf{gibrat} &\\
				& (-0.8182874569530325, & (0.0008101865790087759,\\
				& 4.715577111274514) & 0.6232572274485035)\\
				\hline
5Fuga 			& \textbf{exponential} &\\
				& (6.439843306079128e-05, & (0.000370484070649546,\\
				& 54.80879613615933) & 0.7952495710610652)\\
				\hline
6Cadencia\_pizz 	& \textbf{exponential} &\\
				& (7.006692352852343e-05, & (0.0006698456997479285,\\
				& 19.015521544283985)& 0.7240439027207031)\\
				\hline
Demachy\_Sarabande\_re\_menor & \textbf{exponential} &\\
				& (0.01010064383568212 & (0.0005023489744358511,\\
				& 52.61581092573515) & 0.3447789213387449)\\
				\hline
Demachy\_Suite\_Chaconne\_sol & \textbf{exponential} &\\
				& (0.013249552891189571,& (0.00031424793187317945,\\
				& 66.06544370206078)& 0.6804106066314026)\\
				\hline
Demachy\_Suite\_Gavotte & \textbf{exponential} &\\
				& (0.00907378670237754,& (0.0003542381763386171,\\
				& 43.37149067252313) & 0.8707373245363814)\\
				\hline
Demachy\_suite\_preludio & \textbf{exponential} &\\
				& (0.003703350282242041, & (0.0005173669018155858,\\
				& 36.901870808960744) & 0.892897503373467)\\
				\hline
Demachy\_Suite\_Sarabande\_sol & \textbf{exponential} &\\
				& (0.004814632025998116, & (0.00038845261507602924,\\
				& 51.883852099537926) & 0.5912165235960434)\\
				\hline
Rafa\_Abel\_Arpegiata\_042020 & \textbf{exponential} &\\
				& (0.0024395049012698716,  & (0.0004982473928031861,\\
				& 46.66119893127226) & 0.3533350978512916)\\
				\hline
Sainte\_Colombe\_Chaconne & \textbf{exponential} &\\
				& (0.004781190288554942, & (0.00026155262437843607,\\
				& 77.01942084679918) & 0.7073417842873942)\\
				\hline
Sainte\_Colombe\_tres\_preludios & \textbf{exponential} &\\
				& (0.001360510981818111, & (0.0002800344215738715,\\
				& 72.76295284306076) & 0.3853727912869811)\\
\hline
\end{tabular}
\caption{\label{vbestfit}{\bf Piece of Music, best fit, and KS test.} Probability density function (PDF) of the exponential distribution: $f(x) = exp(-x)$, for $x >= 0$. PDF of the gibrat distribution: $f(x) = \frac{1}{x \sqrt{2\phi}} exp(-\frac{1}{2} (log(x))^2)$, for $x >= 0$.}
\end{table}

\begin{figure}[!h]\centering
	\includegraphics[width=0.85\textwidth]{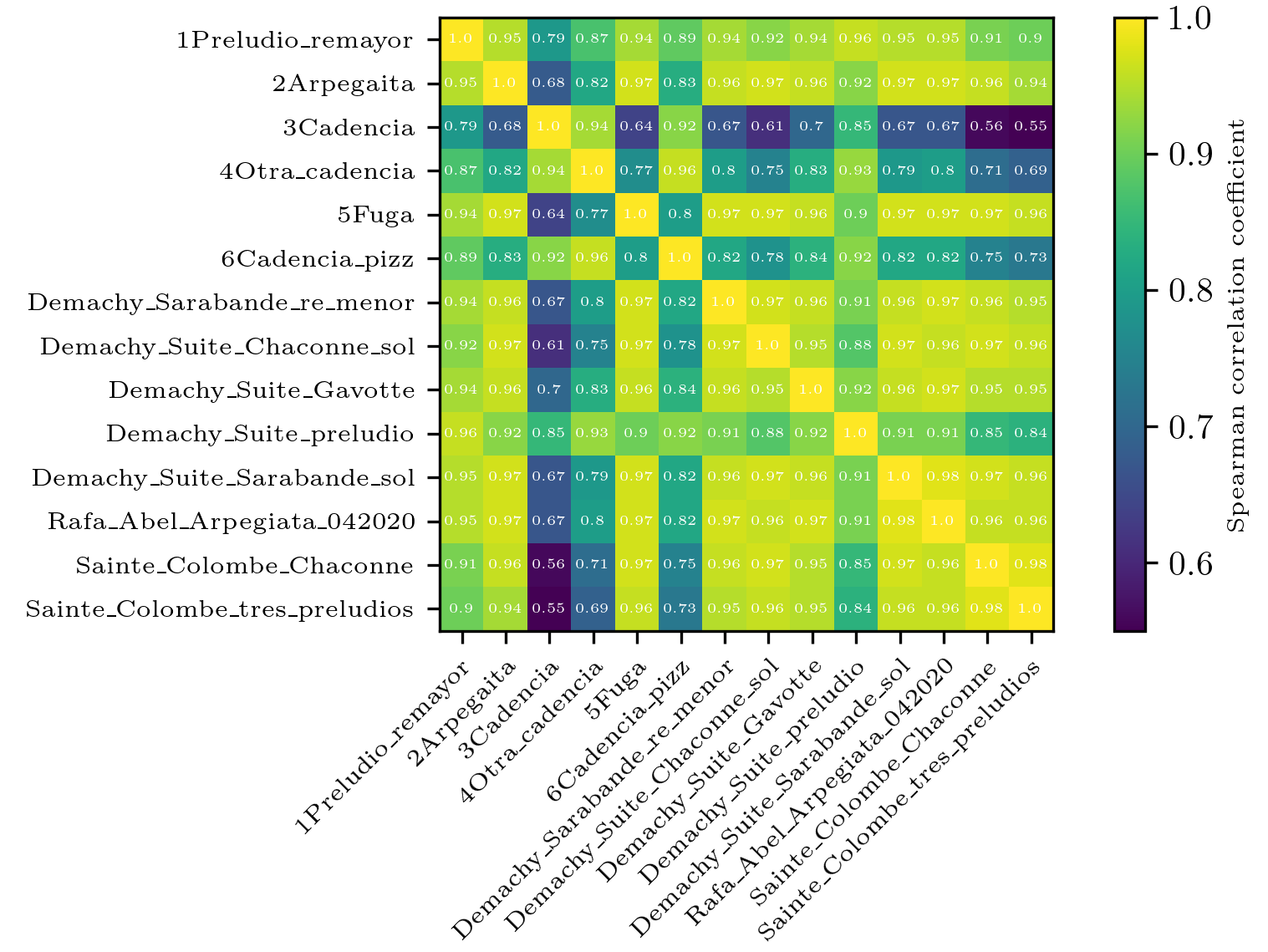}
	{\caption{Degree correlation matrix. We computed the Spearman correlation coefficient and compared with every other musical case \citep{KendallStuart1973}. 
	\label{correlationMatrix}}}
\end{figure}

In the same line as Table \ref{vbestfit}, we can see the underlying behavior related to the large-scale presence of the exponential distribution in most of the cases. Figure \ref{correlationMatrix} presents the degree correlation matrix of the set of pieces of music. Based on the measure of the degree centrality for nodes, in most cases, we can see positive and higher values of correlation when comparing each piece of music to each other. There are three pieces of music that show positive and medium values of correlation when comparing them with each others' pieces of music. These are ``3Cadencia,'' ``4Otra\_cadencia,'' and ``6Cadencia\_pizz.'' Therefore, we can see consistent patterns of correlation values of the degree centrality in 78\% of all cases. 

\begin{figure}[!h]\centering
	\includegraphics[width=0.85\textwidth]{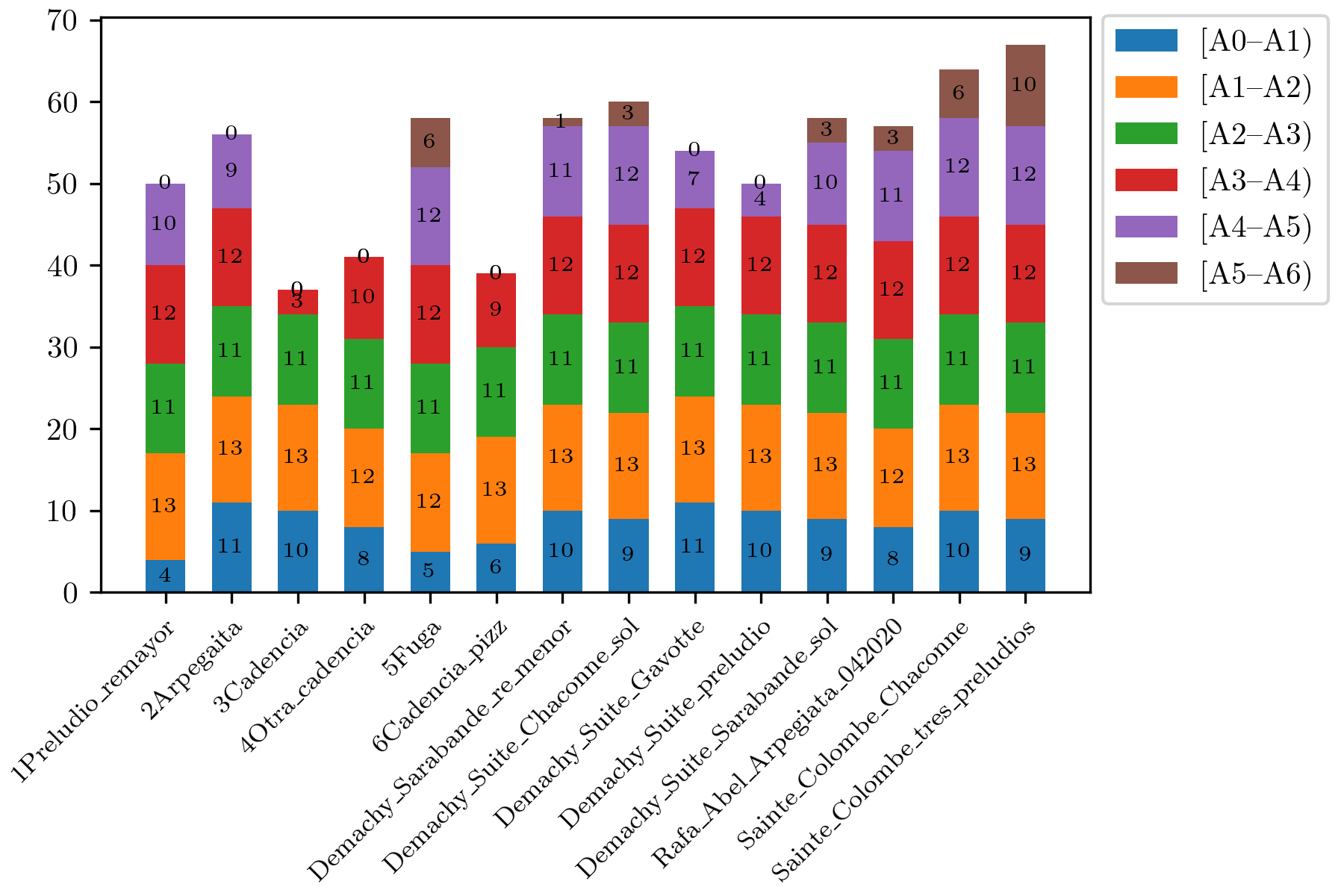}
	{\caption{Number of and type of nodes in the largest cliques
	\label{largestCliques}}}
\end{figure}

Next, Figure \ref{largestCliques} provides a more detail information of similar sounds that are presented in each piece of music. Figure \ref{largestCliques} shows the number and type of sounds related to the largest cliques in each piece of music. As shown in this figure, the number of sounds associated with each largest clique is over $35$ and under $70$. In addition, this figure shows the frequency of sounds that are in most of the ranges of the scientific pitch notation per musical case. Then, all the cases display ranges from $[A0-A1)$ to $[A3-A4)$. However, there are 7 cases (50\%) that do not display sounds of $[A4-A5)$ and $[A5-A6)$. An interesting pattern in this figure is related to the middle frequencies related to $[A1-A2)$, $[A2-A3)$, and $[A3-A4)$ of each case. $[A1-A2)$, $[A2-A3)$ display similar values of frequencies, meanwhile $[A3-A4)$ present two small frequency values that are related to ``4Otra\_cadencia,'' and ``6Cadencia\_pizz.'' Therefore, Figure \ref{largestCliques} shows consistent findings associated with Table \ref{vbestfit} and Figure \ref{correlationMatrix}.

In summary, it has been shown from these results that the statistical regularity related to the presence of the exponential distribution in most of the musical cases is associated with the degree centrality of sounds. This points out the importance of similar sounds through all the cases based on the number of connections to other sounds. These sounds are central points in each network of sounds. Then, each of the largest clique captured in detail some sounds in ranges of sounds that are similar to all cases. Each of the largest clique showed small and significant variations in their number and frequencies in the range of sounds.

\section{Discussion}
This study found that the complex frequency fluctuations can be related to the presence of a singular statistical distribution, which in turn can be described by highly values of degree correlations and different range of sounds in the largest cliques. From this, we can infer that these findings are associated with the interpretative style of a particular musician who plays different pieces of music. The only constants in our musical cases were the same interpreter and his bass viola da gamba. Therefore, the interpreter may imprint particular frequency fluctuations in every piece of music in which predictable pattern of sounds coexists with their unpredictable variations.  

Based on our initial questions, mentioned in the Introduction section, we can state that the network of sounds is a collection and unique realization of related sounds that are closely related to the interpretative style of a particular interpreter. Modeling such a collection of related sounds as a network assists to understand the presence of large-scale patterns and small-scale variations of sounds. Finally, based on the degree centrality correlation matrix and the number of and type of sounds in the largest cliques, we believe that they are fundamental organizing results for analyzing and identifying relationships and variations between sounds. Therefore, modeling a network of sounds based on complex systems and music theory, we can describe objectively the possible relationship between similar patterns of complex frequency fluctuations and the interpretative style of an interpreter in different musical repertoires. 

In the same line as above, it is important to notice the musical cases related to the piece of music of Arpeggiata: audio files of ``2Arpegaita'' and ``Rafa\_Abel\_Arpegiata\_042020.'' These files are two different recordings of the same piece of music, but in different time periods. Such audio files were performed by the same player on the same instrument. Therefore, we found that both cases showed a degree correlation value of $0.97$ and similar number of largest cliques $56$ and $57$ respectively. However, we observed important variations in the number of  sounds related to those largest components. In particular, we showed significant variations in the number of sounds associated with ranges of  $[A0-A1)$, $[A4-A5)$, and $[A5-A6)$. This suggests that such small but significant variations can be related to the technology of recording equipment, the musical venue, and the emotional state of the musician.

We confirm that our hypothesis is correct. We found the presence of a similar statistical distribution---exponential---in the set of different pieces of music due to the performance of a same musician. In addition, we find subtle and significant differences in the complex network measures, Figures \ref{correlationMatrix} and \ref{largestCliques}, because of the granular analysis of a network of sounds. Therefore, we can clearly state that the interpreter imprints a particular frequency fluctuations in every piece of music in which a predictable pattern of sounds coexists with their unpredictable variations.

Overall, these findings may help us to better understanding the following open questions and their possible development pathways in the relationship between the human creativity and the automatization of such a creativity by computer algorithms. Which elements of the sound can generate music? Can any acoustic or sound phenomenon be understood and studied as a musical object?
From the perspective of semiology, music ``cannot be reduced to an unique identity'' \citep{Molino2011}. What we call ``music'' encompasses simultaneously the production of an acoustic material---i.e., the performance---the sound itself, and the reception of that material---i.e., the listening process \citep{Molino2011}. Understanding an event as music, we have to notice the presence of those three aspects, even if they can be asynchronous, as in the case of a recording. Then, the musical performance is ``not a simply reproduction'' of a score, ``but a process experienced in a particular cultural context, created by performers (using the notation) and mentally constructed (uniquely and temporarily) by each listener'' \citep{LeechWilkinson2012}. Therefore, the musical meaning perceived from a performance is intrinsically subjective, which is often very different for the various parts involved in the process.

In addition, following the discussion of the above issues, there is a particularly important open question: Is a recording a musical performance? If we consider an audio file as a digitally encoded computer document, it can store data that is translated as sound when playing on a certain display. That sound might be perceived as music by a listener. However, it does not contain sound or music by itself, but it might be interpreted as that. Therefore, to consider an audio file as music, we need to listen it. We must to remember that an audio file is an encoded data; it is not a musical experience.

In general, we can describe the music as a complex phenomenon that comprises many dimensions of the human activity, such as social, political, and historical activities. In particular, we should not lose sight of the fact that, mixing together the scientific and music approaches, there are different and related analytic perspectives. For example, the analysis of music recordings based on signal processing and perceptual cognitive processes, the theoretical analysis of composition based on the music score, and the analysis of musical performance. Therefore, to study a musical fact, we must focus on its individual and their collective aspects and keep in mind the complexity of the phenomenon and its emergent properties.

Future work should focus on identifying the network of sounds and their characteristics of different musicians across different type of music. In fact, our analysis can be extended beyond the music. This type of analysis can be used in different human arts in which the creativity expression reflects the underlying influence of social and cultural context.

\subsection{Conclusion}
Our main conclusions are the following:
\begin{itemize}
	\item It is not trivial for a musician, who plays a collection of different pieces of music, to imprint a particular frequency fluctuations that are their distinctive and irreducible features of sounds.

	\item There is a clear evidence that the presence of complex frequency fluctuations in a set of different pieces of music playing by the same musician are related to the interpretative style of such a musician. Every musician imprints a particular frequency fluctuations that are their personal and expressive sounds.

	\item Future computer algorithms related to create music should incorporate approximately the sophistication of the human creativity for reproducing complex frequency fluctuations that sound as music. 
\end{itemize}




\end{document}